\documentclass[pre,showpacs,amsmath,amsfonts,endfloats]{revtex4}

\usepackage{bm}

\newtheorem{theorem}{Theorem}
\newtheorem{lemma}{Lemma}

\begin{document}

\title{Geometry and Topology of Escape I: Epistrophes}

\author{K.~A.~Mitchell}
\email{kevinm@physics.wm.edu}
\author{J.~P.~Handley} 
\author{B.~Tighe} 

\affiliation{Department of Physics, College of William and Mary,
Williamsburg, Virginia, 23187-8795}

\author{S.~K.~Knudson} 
\affiliation{Department of Chemistry, College of William and Mary,
Williamsburg, Virginia, 23187-8795}

\author{J.~B.~Delos} 
\email{jbdelo@wm.edu}

\affiliation{Department of Physics, College of William and Mary,
Williamsburg, Virginia, 23187-8795}

\date{\today}

\begin{abstract}
We consider a dynamical system given by an area-preserving map on a
two-dimensional phase plane and consider a one-dimensional line of
initial conditions within this plane.  We record the number of
iterates it takes a trajectory to escape from a bounded region of the
plane as a function along the line of initial conditions, forming an
``escape-time plot''.  For a chaotic system, this plot is in general
not a smooth function, but rather has many singularities at which the
escape time is infinite; these singularities form a complicated
fractal set.  In this article we prove the existence of regular
repeated sequences, called ``epistrophes'', which occur at all levels
of resolution within the escape-time plot.  (The word ``epistrophe''
comes from rhetoric and means ``a repeated ending following a variable
beginning''.)  The epistrophes give the escape-time plot a certain
self-similarity, called ``epistrophic'' self-similarity, which need
not imply either strict or asymptotic self-similarity.
\end{abstract} 

\maketitle

Chaotic transport, and the escape of trajectories from defined regions
of phase space, has been an important topic in dynamics for many
years, because it describes phenomena that occur in many branches of
physics.  For example, some meteorites that fell on Antarctica are
believed to have come from Mars; how they escaped from Mars'
gravitational field is a problem in the theory of chaotic
transport\cite{jaffe02}.  At a smaller scale, one of the important
topics in nanophysics is ballistic transport of electrons through a
small junction: electrons enter a junction from one lead, bounce
around within the junction following either regular or chaotic paths,
and eventually find their way to an exit lead\cite{junctions}.  A
closely related problem is chaotic propagation of light rays in a
distorted cylindrical glass bead\cite{noeckel97}.  At the molecular
level, we may think about the breakup of a temporarily-bound complex,
such as a He atom weakly bound to an $\mbox{I}_2$
molecule\cite{Davis,Tiyapan93,Tiyapan94,Tiyapan95}.  At the atomic
level, the ionization of an excited hydrogen atom in applied electric
and magnetic fields is an ideal candidate for the laboratory study of
chaotic
transport\cite{koch,Jaffe99,Jaffe00,Jaffe01,Wiggins01,Lankhuijzen96,Robicheaux9697}.

We can learn many of the properties of chaotic transport by studying
area-preserving maps of the plane.  We examine the time required to
escape from a specified region of the plane, plotted as a function
along a given line of initial conditions.  Within this escape-time
plot, we study regular sequences of escaping intervals, which we call
``epistrophes''.

\section{Introduction}

We are motivated by the chaotic ionization of a hydrogen atom placed
in strong parallel electric and magnetic fields.  The dynamics of the
hydrogenic electron can be modeled classically by an area-preserving
map on a two-dimensional phase plane.  This map exhibits a prominent
homoclinic tangle (see below), which organizes the dynamics, leading
to phase space transport and eventually escape.  The mechanism of
escape via a tangle is a common model for many classical systems.  In
this paper, we consider the general problem of escape for an arbitrary
map possessing a homoclinic tangle exhibiting the basic structure
shown in Fig.~\ref{fSOSsc}.

The map in Fig.~\ref{fSOSsc} has an unstable fixed point
(\textsf{X}-point) $\mathbf{z}_\mathsf{X}$, with a pair of stable and
unstable manifolds attached to it.  These manifolds are invariant
curves containing all points that asymptote to
$\mathbf{z}_\textsf{X}$ under forward and backward iterates,
respectively \cite{Wiggins92,Jackson90}.  The curves intersect
transversely at the point $\mathbf{P}_0$.  The ``complex'' is the
region bounded by the segments of the stable and unstable manifolds
joining $\mathbf{z}_\textsf{X}$ to $\mathbf{P}_0$; escape is defined
as mapping out of the complex.  As explained by Poincar\'e, the
transverse intersection $\mathbf{P}_0$ produces a homoclinic tangle.
Between successive intersection points $\mathbf{P}_n$ and
$\mathbf{Q}_n$, the manifolds bound lobes denoted $E_n$ and $C_n$ in
Fig.~\ref{fSOSsc}.  As these lobes are mapped forward or backward,
their widths are compressed and their lengths are stretched; they
become long and thin and develop intricate twisted shapes.  The
resulting complex structure of the intersecting stable and unstable
manifolds is called a homoclinic tangle.

An important aspect of any classical decay problem is the distribution
of initial points in phase space.  When modeling the breakup of
molecular collision complexes, for example, one normally assumes that
the complex is more-or-less in thermal equilibrium.  A microcanonical
distribution of initial probability might be used within the collision
complex, with equal probabilities in equal areas, or perhaps some
other smooth distribution.  However, in most experiments on excited
atoms in strong fields, the initial distribution is quite different.
The electron attains a high energy by single-photon excitation from a
localized strongly-bound initial state \cite{welge,
walther,kleppner,Amsterdam,Eichmann88,Spellmeyer97}.  The trajectories
therefore start close to the nucleus and go out in all directions.
This initial distribution is well modeled by assuming that all
electrons begin exactly at the nucleus, with constant energy, and with
a smooth distribution of outgoing directions \cite{Du88,Gao92,Main94}.
In the phase plane, this yields a distribution of initial states along
a {\em line} of initial conditions.  Thus, we assume that the initial
distribution of states in phase space lies along some curve
$\mathcal{L}_0$, the details of which depend on the problem at hand.
(See Fig.~\ref{fSOSsc}.)  Along this curve there is some initial
density of points; for hydrogen this density represents the initial
angular distribution of outgoing electrons \cite{footnote1} .

We plot the number of iterates $n_i$ needed to escape as a function
along the line of initial conditions $\mathcal{L}_0$, forming an {\em
escape-time plot}, as shown in Fig.~\ref{fEPSC}.  Each line segment in
Fig.~\ref{fEPSC} represents an interval, or \textit{escape segment},
of $\mathcal{L}_0$, in which all points escape the complex at the same
iterate.  The escape-time plot is clearly a very complicated function
with ``fractal'' properties; this fractal structure is created by the
repeated intersections of the stable manifold with the line of initial
conditions.  Our objective is to describe certain regular structures
within this plot.

Figure~\ref{fEPSC} contains many prominent sequences of escape
segments, several of which are indicated by bold arrows.  We call each
such sequence an \textit{epistrophe}.  Epistrophes have several
important properties.  (1) Beginning at some initial iterate, each
epistrophe contains one escape segment at every subsequent iterate.
(2) Each epistrophe converges to some point on $\mathcal{L}_0$.  (3)
Within a given epistrophe, the lengths of the escape segments decrease
geometrically (in the limit $n_i \rightarrow \infty$) with the ratio
of successive lengths converging to the Liapunov factor (i.e. the
largest eigenvalue) $\alpha$ of the \textsf{X}-point.  This is true
regardless of which epistrophe we analyze. 

The epistrophes form hierarchical sequences -- we see in
Fig.~\ref{fEPSC} that the endpoints of each escape segment serve as
the limit points for epistrophes beginning at higher iterate number.
For example, consider epistrophe \textsf{a}, which begins at $n_i = 5$
(around $p = 0.56$) and progresses upward, containing segments at $n_i
= 6, 7, 8, ...$ .  Upon each of the two endpoints of the first segment
($n_i = 5$), there converges another epistrophe which begins at $n_i =
11$.  Similarly, the second segment of epistrophe \textsf{a} ($n_i =
6$) has an epistrophe converging upon each of its endpoints, beginning
at $n_i = 12$.  In fact, every escape segment has an epistrophe which
converges upon each of its endpoints.  Thus, epistrophes appear
throughout the escape-time plot and on all scales.

The main result of this paper is the Epistrophe Theorem
(Sect.~\ref{sEpistrophes}), which proves and elaborates upon the above
observations for an arbitrary homoclinic tangle and an arbitrary line
of initial conditions.  The beginning of each epistrophe is not
described by the Epistrophe Theorem, leaving a certain
unpredictability in how an epistrophe starts.  What is described is
the asymptotic behavior of the tail of the epistrophe.  In fact, we
prove that, up to an overall rescaling, the asymptotic tails of all
epistrophes are identical; we characterize these tails with geometric
quantities ($\alpha$, $\chi$, and $\phi$ in Theorem \ref{t5}).

The recursive nature of the Epistrophe Theorem and the scaling
relation between the epistrophe tails implies a certain
self-similarity to the escape-time plot.  However, the Epistrophe
Theorem is itself not strong enough to imply true self-similarity (or
even asymptotic self-similarity).  Our data indicate that there may be
numerous escape segments, which we call ``strophes'', that do not
belong to {\em any} epistrophe and that even tend to dominate the
escape-time plot at long times.  Several such strophes are indicated
by asterisks in Fig.~\ref{fEPSC}.  We can thus say that the Epistrophe
Theorem implies a kind of ``epistrophic'' self-similarity: epistrophes
(self-similar sequences) occur on all scales, but there may also be
additional segments, or strophes, that persist (and may even dominate)
in the asymptotic limit.

On the left-hand side of Fig.~\ref{fEPSC} is plotted the winding
number $n_w$ of escape, i.e. the number of times a trajectory winds
around the ``center'' of the complex as it escapes to infinity.  (In
this case the center is the stable zone in Fig.~\ref{fSOSsc}.)  The
data show that the winding number $n_w$ is constant along an
epistrophe.  For example, all escape segments in epistrophe \textsf{a}
have $n_w = 1$; all segments in epistrophe \textsf{d} have $n_w =
2.5$; and so on.  Also, the winding number of an epistrophe is always
one greater than the winding number of the segment upon which it
converges.  For example, epistrophe \textsf{c} has $n_w = 2$ and
converges upon a segment of \textsf{a} with $n_w = 1$.  In a separate
publication, we will prove several theorems explaining these
observations.

The impact of tangles on chaotic transport has been an active field of
research for at least the last twenty years, with notable
contributions by MacKay, Meiss, and Percival \cite{MacKay84}, Davis
and Gray \cite{Davis}, Rom-Kedar \cite{Rom-Kedar90,Rom-Kedar94},
Wiggins \cite{Wiggins92}, and numerous others.  More specifically,
there has been significant interest in the fractal behavior of
escape-time plots (or alternatively, scattering functions) in a
variety of fields, including work by Noid et al. \cite{Noid86}, Petit
and Henon \cite{Petit86}, Eckhardt \cite{Eckhardt86,Eckhardt87}, Jung
and coworkers
\cite{Jung87,Ruckerl94a,Ruckerl94b,Lipp95,Jung99,Butikofer00}, and
Gaspard and coworkers \cite{Gaspard89,Gaspard98}.

Our research was inspired by the work of Tiyapan and Jaff\'e on the
scattering of He from an excited $\text{I}_2$ dimer
\cite{Tiyapan93,Tiyapan94,Tiyapan95}.  In their study (particularly
Ref.~\cite{Tiyapan93}), Tiyapan and Jaff\'e examined a final-action
versus initial-angle plot (analogous to the escape-time plot) over a
wide range of scales.  In their numerical data, they identified
infinite sequences equivalent to our epistrophes.  Jung and coworkers
have also extensively studied scattering functions.  In particular,
they partially label the asymptotically bound orbits using a symbolic
dynamics, which captures important topological structures of the
scattering functions.  As in the work of Tiyapan and Jaff\'e, the
sequences which we call epistrophes are present in their description.
However, none of the above work gives a full characterization of the
epistrophes, nor does it give a proof that all epistrophes of a given
map are asymptotically self-similar and asymptotically similar to each
other.  This is the primary result of the present paper and is
summarized in the Epistrophe Theorem.  The Epistrophe Theorem is also
closely related to (but distinct from) Palis's ``$\lambda$-lemma''
\cite{Palis69,footnote2}.

Our paper has the following structure.  Section~\ref{sSaddleCenter}
states the technical assumptions we require of the tangle.  Section
\ref{sEpistrophes} contains the Epistrophe Theorem.  Implications for
the fractal structure of the escape-time plots are discussed in
Sects.~\ref{sEtymology} -- \ref{sFractals}.  Section~\ref{sDiscussion}
contains conclusions and a discussion of future work.  Appendix
\ref{sSCTheory} defines our example map.  The proof of the Epistrophe
Theorem is contained in Appendix \ref{sEpistropheProof}.

\section{Homoclinic Tangles}

\label{sSaddleCenter}

\subsection{Assumptions on the Map}

\label{sAssumptions}

We consider an arbitrary ``saddle-center map'' $\mathcal{M}$ which has
a simple homoclinic tangle (as shown in Fig.~\ref{fSOSsc}) that is
described by the following five assumptions.

\noindent \textbf{Assumption 1} \textit{The map $\mathcal{M}$ is a
canonical map or, more precisely, an analytic area- and
orientation-preserving diffeomorphism of an open subset of the phase
plane.}

\noindent \textbf{Assumption 2} \textit{The map has an unstable fixed
point (\textsf{X}-point) $\mathbf{z}_{\textsf{X}}$, without
inversion.}

\noindent \textbf{Assumption 3} \textit{Proceeding away from the
\textsf{X}-point, one branch of the stable manifold and one
branch of the unstable manifold (called the nontangled manifolds) each
go to infinity without intersecting any other stable or unstable
manifold; the other branch $\mathcal{S}$ of the stable manifold and
$\mathcal{U}$ of the unstable manifold (called the tangled manifolds)
intersect each other transversely.}

Assumption 3 is essentially Rom-Kedar's definition of an ``open
map''\cite{Rom-Kedar94}.  A primary intersection point, or ``pip'',
$\mathbf{z}_{pip}$ is a transverse intersection between $\mathcal{S}$
and $\mathcal{U}$ such that the segment of $\mathcal{S}$ joining
$\mathbf{z}_\textsf{X}$ to $\mathbf{z}_{pip}$ does not intersect the
segment of $\mathcal{U}$ joining $\mathbf{z}_\textsf{X}$ to
$\mathbf{z}_{pip}$\cite{Easton86,Wiggins92}.  We choose a pip
$\mathbf{P}_0$ such that $\mathcal{U}$ crosses $\mathcal{S}$ from
right to left.  The \textit{complex} is defined as the region enclosed
by $\mathcal{S}$ and $\mathcal{U}$ from $\mathbf{z}_\textsf{X}$ to
$\mathbf{P}_0$; it contains its boundary, including
$\mathbf{z}_\textsf{X}$.  The forward and backward iterates of
$\mathbf{P}_0$ are also homoclinic intersections (with the same sense)
and are denoted $\mathbf{P}_n = \mathcal{M}^n(\mathbf{P}_0)$, $-\infty
< n < \infty$.

\noindent \textbf{Assumption 4} \textit{Between $\mathbf{P}_0$ and
$\mathbf{P}_1$, $\mathcal{S}$ and $\mathcal{U}$ intersect just once, at a
point we call $\mathbf{Q}_{0}$.}

The intersection $\mathbf{Q}_0$ has the opposite sense as
$\mathbf{P}_0$ as do its forward and backward iterates $\mathbf{Q}_n =
\mathcal{M}^n(\mathbf{Q}_0)$, $-\infty < n < \infty$.  The segments of
$\mathcal{S}$ and $\mathcal{U}$ between $\mathbf{P}_0$ and
$\mathbf{Q}_0$ enclose the {\em escape zone} $E_0$, which by
definition contains its outer $\mathcal{U}$-boundary but not its inner
$\mathcal{S}$-boundary (and neither $\mathbf{P}_0$ nor
$\mathbf{Q}_0$.)  Similarly, the segments of $\mathcal{S}$ and
$\mathcal{U}$ between $\mathbf{Q}_{-1}$ and $\mathbf{P}_0$ enclose the
{\em capture zone} $C_0$, which by definition contains its outer
$\mathcal{S}$-boundary but not its inner $\mathcal{U}$-boundary (and
neither $\mathbf{P}_0$ nor $\mathbf{Q}_{-1}$.)  The forward and
backward iterates of $E_0$ and $C_0$ are called the \textit{escape
zones} $E_n$ and the \textit{capture zones} $C_n$ ($-\infty < n <
\infty$).

The lobes $C_0$ and $E_{-1}$ form a \textit{turnstile}
\cite{MacKay84,Wiggins92}: on one iterate of the map, all points in
$C_0$ map into the complex, i.e. are captured, and all points in
$E_{-1}$ map out of the complex, i.e. escape.  It is important to
recognize that all points which eventually escape the complex lie in
some escape zone $E_{-k} = \mathcal{M}^{-k}(E_0)$, $k > 0$.

\noindent \textbf{Assumption 5} \textit{Mapping forward causes all
points in $E_0$ to march off to infinity, never to re-enter the
complex.  Likewise mapping backward causes all points in $C_0$ to
march off to infinity, never to re-enter.}

Assumption 5 means that no point can escape from and subsequently
return to the complex; equivalently, $E_n \cap C_{n'} = \emptyset$ for
$n \ge 0$ and $n' \le 0$.

By convention we orient the tangle in the $pq$-plane as shown in
Figs.~\ref{fSOSsc} and \ref{fX1}; that is, $\mathbf{P}_0$ is west of
$\mathbf{z}_\mathsf{X}$, and $\mathcal{S}$ is north of $\mathcal{U}$ when
linearized about $\mathbf{z}_\mathsf{X}$.

\subsection{Canonical Length on the Stable and Unstable Manifolds}

For any point $\mathbf{z}_0 \in \mathcal{S}$, there is a natural
length $s(\mathbf{z}_0)$ along the stable manifold, as measured from
$\mathbf{z}_\textsf{X}$.  Setting $\mathbf{z}_n =
\mathcal{M}^n(\mathbf{z}_0)$, we define
\begin{equation}
s(\mathbf{z}_0) 
= \lim_{n \rightarrow \infty} |\mathbf{z}_n - \mathbf{z}_\textsf{X} | \alpha^n,
\label{r8}
\end{equation}
where $| \; \; |$ is the standard Euclidean vector norm.  The limit in
Eq.~(\ref{r8}) is well-defined since after each iterate,
$|\mathbf{z}_n - \mathbf{z}_\textsf{X} |$ decreases by a factor
$\alpha$ within the linear approximation to $\mathcal{M}$.  It follows
from Eq.~(\ref{r8}) that
\begin{equation}
s(\mathbf{z}_0)
= s(\mathbf{z}_n) \alpha^n.
\label{r42}
\end{equation}
Up to a constant scale factor, Eq.~(\ref{r42}) uniquely determines
$s$.  Under a canonical transformation, the function $s(\mathbf{z})$
only changes by an overall scale factor.  Hence, we call $s$ the {\em
canonical length} along $\mathcal{S}$, and we call $|s(\mathbf{z}) -
s(\mathbf{z}')|$ the canonical length between $\mathbf{z}$ and
$\mathbf{z}'$.

Analogous to $s$, we define the canonical length $u$ along
$\mathcal{U}$ by
\begin{equation}
u(\mathbf{z}_0) 
= \lim_{n \rightarrow \infty} |\mathbf{z}_{-n} - \mathbf{z}_\textsf{X}| 
\alpha^n,
\label{r10}
\end{equation}
which satisfies
\begin{equation}
u(\mathbf{z}_0)
= u(\mathbf{z}_{-n}) \alpha^n.
\end{equation}

Recall that the lobes $E_n$ and $C_n$ each have a
$\mathcal{U}$-boundary and an $\mathcal{S}$-boundary.  We denote the
canonical lengths of these boundaries by
\begin{subequations}
\begin{eqnarray}
u(E_0) & = & |u(\mathbf{Q}_0) - u(\mathbf{P}_0)|, 
\label{r55} \\
s(E_0) & = & |s(\mathbf{Q}_0) - s(\mathbf{P}_0)|, \\
u(C_0) & = & |u(\mathbf{P}_0) - u(\mathbf{Q}_{-1})|, 
\label{r56} \\
s(C_0) & = & |s(\mathbf{P}_0) - s(\mathbf{Q}_{-1})|.
\label{r11}
\end{eqnarray}
\end{subequations}

\subsection{The Curve of Initial Conditions}

We choose a (differentiable) curve of initial conditions
$\mathcal{L}_0$ that passes through the complex.  We introduce on
$\mathcal{L}_0$ the Euclidean line element $d\lambda = (dp^2 +
dq^2)^{1/2}$, which gives us a coordinate $\lambda$ on $\mathcal{L}_0$
and a length $\lambda_{ab} = | \lambda_a - \lambda_b |$ of the segment
between endpoints $\lambda = \lambda_a$ and $\lambda = \lambda_b$.  We
allow for an arbitrary positive density $d\mu = \rho(\lambda)
d\lambda$ of initial points on $\mathcal{L}_0$, which defines the
measure
\begin{equation}
\mu_{ab} 
= \left|\int_{\lambda_a}^{\lambda_b} \rho(\lambda)d\lambda \right|.
\label{r46}
\end{equation}

\section{Epistrophes and Epistrophic Fractals}

\label{sEpistrophes}

\subsection{The Epistrophe Theorem}

\label{sVIA}

A segment of the line of initial conditions $\mathcal{L}_0$ that
escapes the complex in $k$ iterates lies in the intersection of
$\mathcal{L}_0$ with the escape zone $E_{-k}$.  It is, in fact, one
connected component of this intersection.  With rare exceptions, the
endpoints of an escape segment are therefore points on the stable
manifold $\mathcal{S}$ \cite{footnote3}.  We will prove that upon any
(transverse) intersection between $\mathcal{S}$ and $\mathcal{L}_0$,
there converges an epistrophe of escape segments.

The existence of epistrophes is suggested by the following argument.
In a small neighborhood of $\mathbf{z}_\textsf{X}$, the map
$\mathcal{M}$ is almost linear and can be re-expressed in new
canonical coordinates $(Q(q,p), P(q,p))$ by $(Q,P) \mapsto (Q',P')$,
where \cite{footnote4}
\begin{subequations}
\begin{eqnarray}
Q' & = & \alpha Q + O(2), \\
P' & = & (1/\alpha) P + O(2).
\end{eqnarray}
\label{r51}
\end{subequations}
Ignoring the higher order terms, $\mathcal{S}$ and $\mathcal{U}$ are
respectively the positive $P$ axis and negative $Q$ axis.  Suppose
$\mathcal{L}_0$ is a horizontal line intersecting the positive $P$
axis near $\mathbf{z}_\textsf{X}$, as shown in Fig.~\ref{flinear}.
Then, again ignoring higher order terms, the width $u(E_{-k})$ of the
base of an escape zone decreases by a factor of $\alpha$ on each
backward iterate.  Thus, as the escape zone is mapped backward, it is
squeezed by $1/\alpha$ in the $Q$ direction and stretched by $\alpha$
in the $P$ direction.  Clearly, these lobes must eventually intersect
$\mathcal{L}_0$, and their intersections $\epsilon_k$ must form an
infinite sequence of geometrically decreasing intervals, converging
upon the intersection $\mathbf{z}_{\mathcal{S}}$ of $\mathcal{L}_0$
with $\mathcal{S}$; the Euclidean lengths $\lambda_k$ decrease as
$\lambda_{k+1} = \lambda_k/\alpha$ (in the limit $k \rightarrow
\infty$.)

The Epistrophe Theorem asserts that such geometric sequences appear
for any map with a homoclinic tangle (satisfying Assertions 1 -- 5)
and for any differentiable curve $\mathcal{L}_0$ intersecting the
stable manifold $\mathcal{S}$ at any distance from 
$\mathbf{z}_\textsf{X}$, even far from the region where the
linearization (\ref{r51}) is sensible.

\begin{theorem}[Epistrophe Theorem]  
\label{t5}
Let $\mathcal{M}$ be any ``saddle-center map'' (as defined by
Assumptions 1--5) and $\mathbf{z}_{\mathcal{S}}$ be any transverse
intersection between the stable manifold $\mathcal{S}$ and the
differentiable curve of initial conditions $\mathcal{L}_0$.  For each
$k>0$ choose the escape segment $\epsilon_k \subset \mathcal{L}_0 \cap
E_{-k}$ closest to $\mathbf{z}_{\mathcal{S}}$ (as measured along
$\mathcal{L}_0$.)  Then there is some initial $k_0$ such that for all
$k\ge k_0$, the escape segment $\epsilon_k$ exists and:

\noindent (i) The segments $\epsilon_k$ converge monotonically upon
$\mathbf{z}_{\mathcal{S}}$ (i.e. the distance between $\epsilon_k$ and
$\mathbf{z}_{\mathcal{S}}$ decreases monotonically.)

\noindent (ii) Define: $\mu_k$ -- the measure of $\epsilon_k$ [using
Eq.~(\ref{r46})]; $\gamma_k$ -- the measure between $\epsilon_k$ and
$\epsilon_{k+1}$; $\delta_k$ -- the measure between $\epsilon_k$ and
$\mathbf{z}_{\mathcal{S}}$.  Provided $\rho(\mathbf{z}_{\mathcal{S}})
\ne 0$, all three measures converge geometrically to zero as
\begin{subequations}
\begin{eqnarray}
\lim_{k \rightarrow \infty} \mu_k \alpha^k & = & K_\mu > 0, 
\label{r47} \\
\lim_{k \rightarrow \infty} \gamma_k \alpha^k & = & K_\gamma > 0, 
\label{r48} \\
\lim_{k \rightarrow \infty} \delta_k \alpha^k & = & K_\delta > 0,
\label{r49} 
\end{eqnarray}                   
\end{subequations}
where $\alpha$ is the Liapunov factor of $\mathbf{z}_\mathsf{X}$ and
$K_\mu$, $K_\gamma$, $K_\delta$ are positive real numbers.
Furthermore,
\begin{subequations}
\begin{eqnarray}
\lim_{k \rightarrow \infty} {\mu_k \over \gamma_k} 
= {K_\mu \over K_\gamma} 
& \equiv & \chi > 0,
\label{r50} \\
\lim_{k \rightarrow \infty} {\mu_k \over \delta_k} 
= {K_\mu \over K_\delta} 
& \equiv & \phi > 0,
\label{r57}
\end{eqnarray}
\end{subequations}
where $\chi = u(E_{0})/u(C_{0})$ and $\phi = u(E_{0})/u(\mathbf{P}_0)
= \chi (\alpha -1)/(\alpha + \chi)$.  [The lengths $u(\mathbf{P}_0)$,
$u(E_{0})$ and $u(C_{0})$ are defined in Eqs.~(\ref{r10}), (\ref{r55})
and (\ref{r56}).]
\end{theorem}

The constants $K_\mu$, $K_\gamma$, $K_\delta$ depend upon which
epistrophe is examined, but their ratios do not, as evident from the
formulas for $\chi$ and $\phi$.  This result was unexpected to us; we
were struck by the fact that the same factor $\alpha$ \textit{and} the
same ratios $\chi$ and $\phi$ apply to each epistrophe, no matter how
far it is from $\mathbf{z}_\mathsf{X}$.

The Epistrophe Theorem describes the tails of sequences.  The value of
$k_0$ and the exact values of $\mu_k$, $\gamma_k$, and $\delta_k$
(especially the early values) cannot be predicted from the present
considerations.  What can be predicted is how $\mu_k$, $\gamma_k$, and
$\delta_k$ decay in the asymptotic limit.

The Epistrophe Theorem is proved in Appendix~\ref{sEpistropheProof}.
Equations (\ref{r47}), (\ref{r48}) and Eq.~(\ref{r50}) are verified
numerically in Figs.~\ref{fmu}, \ref{fgamma}, and \ref{fchi}.  These
plots indicate that the asymptotic behavior predicted by the theorem
is approached quickly.

Notice that no epistrophe converges upon the boundary of a stable
zone.  For example, in the upper half of Fig.~\ref{fEPSC}, epistrophe
\textsf{a} converges upward to the boundary of the complex, but there
is no epistrophe converging downward where the boundary of the stable
zone lies.  Similarly, in the bottom half of Fig.~\ref{fEPSC},
epistrophe \textsf{b} converges downward to the boundary of the
complex, but no epistrophe converges upward to the stable zone.

\subsection{\label{sEtymology} Epistrophes and Strophes}

Any orderly infinite sequence of escape segments predicted by the
Epistrophe Theorem is called an \textit{epistrophe}.  The irregular
beginning of such a sequence, or any escape segment or group of
segments that is not part of an epistrophe, is called a
\textit{strophe}.  The strophes contain the unpredicted behavior in
the escape-time plot.

It is helpful to examine the origin and meaning of the word
``epistrophe'' because the parts of this word describe the structure
of epistrophic fractals.  One dictionary \cite{dicta} defines
``strophe'' simply as a stanza of a poem or ballad, while another
\cite{dictb} defines it as a stanza that might have irregular
structure, such as variable length and rhythm.  This ambiguity is
useful to us, because we might or might not find regular structure in
the escape segments that we call strophes.

``Epi-'' is used here in the sense of ``the end of'' or ``concluding''
(as in epidermis or epilogue.)  An epistrophe in rhetoric is a
repeated ending following a variable beginning.  One of the most
familiar in American English is from Lincoln's Gettysburg Address:
``of the people, by the people, and for the people''.  Here ``of'',
``by'', and ``for'' are the strophes and the repetitions of ``the
people'' are epistrophes.  This is an example in which the epistrophes
dominate the structure, as the strophes each have one syllable, while
the epistrophes have three.  Analogous behavior may occur for
dynamical epistrophes: if the Liapunov factor $\alpha$ is close to
one, the total length of the segments in an epistrophe will tend to
dominate over the length in the strophe.

A quite different epistrophic structure is contained in the Hebrew
creation recitative, with its description of the first six days.  The
descriptions of each day (the strophes) vary in length and structure.
Each strophe ends with the epistrophe: ``And the evening and the
morning were the [$n$th] day.'' $n = 1, ..., 6$.  The epistrophe is
short, and the strophes dominate the length of the narrative.  This
tends to happen in dynamical epistrophes if the Liapunov factor
$\alpha$ is large.

Actually the creation recitative might be called a doubly-epistrophic
narrative, because it contains seven repetitions of a different
epistrophe: ``...~and God saw that it was good.''  These are
interspersed with a rhythmic structure different from that of the
``$n$th day'' epistrophe; the second day does not contain this
epistrophe, but it appears in other days once or twice, sometimes at
the end of the day and sometimes in the middle.  Complicated
interleaving of two or more different families of epistrophes
(typically with different $\alpha$) can occur, for example, in a
dynamical system if the boundary of the complex contains more than one
\textsf{X}-point.

For dynamical applications, we define an ``epistrophe'' as an infinite
sequence of escape segments having the properties described by the
Epistrophe Theorem; consistent with rhetoric, an epistrophe has a
predictable ending.  We define ``strophe'' less precisely, consistent
with its use in rhetoric, and apply it in two contexts: (1) A strophe
is the unpredicted beginning of an epistrophe; that is, the first few
escape segments $\epsilon_{k_0}$, $\epsilon_{k_0+1}$, $...$ of an
epistrophe (with unpredicted values of $k_0$, $\mu_{k_0}$, etc.)
constitute one strophe.  In the following paper, we will show that
there is partial predictability of these strophes.  (See also
Sects.~\ref{sSelfSimilarity} and \ref{sDiscussion}.)  (2) A strophe is
any additional escape segment, or group of escape segments, that is
not part of an epistrophe.  Several such strophes were indicated by
asterisks in Fig.~\ref{fEPSC}.  For now, we deliberately avoid giving
the word ``strophe'' a sharp definition; we leave open the possibility
that the strophes might later be described in some more complex
framework.

\subsection{Epistrophic Self-Similarity}

\label{sSelfSimilarity}

The Epistrophe Theorem implies a certain recursive structure to the
escape-time plot: every epistrophe is asymptotically self-similar and
is asymptotically related to any other epistrophe by a change in
scale.  In this sense, there is a kind of asymptotic self-similarity
of sequences of escape segments in the escape-time plot.  A second
type of regularity will be established in the next paper, where we
will state and prove an ``Epistrophe Start Rule''.  As a result of the
global topology of the tangle, there is a ``minimal set'' of escape
segments.  In this minimal set, an epistrophe begins at an iterate
$k_0$, exactly $\Delta = D+1$ iterates after the segment upon which it
converges.  (We will explain that the parameter $D$ is the minimal
delay time of the tangle.)  Hence, in the minimal set there is a
simple recursive pattern to the escape segments: on each side of a
given escape segment, a new epistrophe begins $\Delta$ iterates later,
and these epistrophes converge to the given escape segment as
described by the Epistrophe Theorem; then on each side of each segment
in the new epistrophes, a further new epistrophe begins $\Delta$
iterates later; at every level, every segment spawns two new
epistrophes, {\em ad infinitum}.  One might expect this to produce a
regular self-similar (or at least asymptotically self-similar) fractal
structure.  Indeed, this is readily seen in the ``standard'' Smale
horseshoe (e.g. Ref.~\cite{Jung99}).

Strict self-similarity and asymptotic self-similarity are consistent
with the Epistrophe Theorem and the Epistrophe Start Rule, but they
are not guaranteed by these results.  First, these results allow the
beginnings of the epistrophes to contain irregular lengths that do not
follow any simple pattern.  Indeed in our numerical studies we do not
see any simple pattern to the lengths of the first segment of an
epistrophe.  More importantly, we find additional, unpredicted strophe
segments which are not part of the minimal set, and which therefore do
not fit the pattern of the minimal set.  Furthermore, numerical
evidence seems to indicate that these strophe segments tend to
dominate at long times.

It is possible, and even likely, that the strophes obey some
higher-order and more complex recursive rules.  If one could uncover
these rules, one might hope to find a deeper and more complex kind of
asymptotic self-similarity in the escape-time plot; one might hope
that there would be a finite number of such rules.  However, it is
generally acknowledged
\cite{Easton86,Rom-Kedar90,Rom-Kedar94,Ruckerl94a} that to describe
the topology of a tangle requires a countable infinity of topological
parameters, reflecting the growing topological complexity of the
tangle on finer and finer scales.  This situation is nicely described
by R\"uckerl and Jung \cite{Ruckerl94a}.  Thus, for any finite
description of the tangle, one expects eventually to discover in the
escape-time plot additional structure which had not yet been predicted
and which eventually comes to dominate the structure that is
predicted.

We use the term ``epistrophic self-similarity'' to describe the above
situation: throughout the escape-time plot and on all scales there are
epistrophes; they are all asymptotically self-similar and each is
similar to every other.  However, there may also be unpredicted
strophe segments which also occur on all scales and which may come
to dominate the regular epistrophe structure.

What is the distinction between asymptotic and epistrophic
self-similarity?  A fractal has asymptotic self-similarity if through
repeated magnifications about any point of the fractal, the pattern
converges to an asymptotic structure.  Epistrophic self-similarity is
weaker, requiring only the existence of epistrophes, as described
above.  If when repeatedly magnifying the fractal, one continues to
see new structures emerge, then the fractal certainly does not possess
asymptotic self-similarity, but may still possess epistrophic
self-similarity.

\subsection{Epistrophic Fractals}

\label{sFractals}

Thus far, we have focused on the segments of $\mathcal{L}_0$ that
escape the complex.  The question remains: How do the epistrophes
influence the points that survive and never escape?  Jung and Scholz
observed that the stable manifolds of all the bound orbits give rise
to a Cantor set of singularities for a scattering function
\cite{Jung87}.  We show that the Epistrophe Theorem itself directly
leads to a Cantor set within the set of surviving points.

Obviously any point inside a stable island survives.  Accordingly, we
define the \textit{stable domain} $S$ to be the union of all open
disks each of which contains no escaping points.  The set $S$ is an
open set in the plane, containing the interiors of all the stable
islands.  Any point in $S$ is \textit{stably surviving}, in the sense
that any sufficiently small displacement of the point still results in
survival.  (By definition, $S$ contains all stably surviving points.)
Continuity guarantees that points on the boundary of $S$ (the
``shoreline'' of the stable islands) also survive.  However, they are
\textit{unstably surviving} in the sense that a small perturbation
can cause them to escape.

There are other sets of unstably surviving points in the complex
besides the shoreline of the islands of stability.  (1) The
\textsf{X}-point $\mathbf{z}_\textsf{X}$ is unstably surviving.  (2)
Many unstable periodic orbits are embedded in the chaotic sea
surrounding the islands of stability.  (3) There may be entire curves
of neutrally stable periodic orbits.  (4) Each unstable periodic orbit
has stable manifolds which do not escape.  (5) There may be Cantori of
surviving points in the chaotic sea; a Cantorus is an invariant Cantor
set which is the remnant of a dissolved KAM torus \cite{MacKay84}.
(6) There are chaotic trajectories that wander about in the chaotic
sea but never escape.  There might be other types of unstably
surviving points which we have omitted here.

The line of initial conditions $\mathcal{L}_0$ runs through the
complex and may intersect some of the above-mentioned sets of
surviving points, while missing others.  Certainly, $\mathcal{L}_0$
intersects the stable manifold of $\mathbf{z}_\textsf{X}$ an infinite
number of times.  We also expect that $\mathcal{L}_0$ may intersect
the set $S$ of stable islands.  Each such intersection will produce an
open segment of $\mathcal{L}_0$ which does not escape, bounded by
surviving endpoints which are part of the shoreline of $S$.  These are
the \textit{only} generic intervals of $\mathcal{L}_0$ known to us
that do not escape, and we often simply assume this case
\cite{footnote5}.

The Epistrophe Theorem yields specific results about the structure of
the set $\mathbb{S}_T$ of surviving points of $\mathcal{L}_0$.  The
set of escaping points is denoted $\mathbb{E}$.  (See
Fig.~\ref{fsets}.)  Since each escape segment is open, the set
$\mathbb{E}$ is also open \cite{footnote6}.  The set $\mathbb{S}_T$,
being everything not in $\mathbb{E}$, is closed (and hence compact.)

We define $\mathbb{S}$ to be the interior of the set $\mathbb{S}_T$;
that is, $\mathbb{S}$ is the union of all open sets in $\mathbb{S}_T$.
As noted above, generically $\mathbb{S}$ consists entirely of points
in the interior of stable islands.  Any surviving point not in
$\mathbb{S}$ we put in a set denoted $\mathbb{F}$, so $\mathbb{F}$
contains whatever points of $\mathcal{L}_0$ survive, other than open
intervals; all points in $\mathbb{F}$ are unstably surviving.  The set
$\mathbb{F}$ is constructed by the following process.  After a finite
number $n$ of iterates of the map, a finite number of escape segments
will have been removed (assuming that ${\cal L}_0$ is analytic.)
Between these escape segments are intervals that have managed to
survive for $n$ iterates.  As we continue to iterate the map, we
remove subintervals of these surviving intervals.  In the limit, if
there are any surviving intervals, we put them into the set
$\mathbb{S}$.  What remains is $\mathbb{F}$.  This is very much like
the construction of the Cantor middle-third set.  In fact, we
demonstrate below that $\mathbb{F}$ is a \textit{topological} Cantor
set.

A topological Cantor set is any set which is homeomorphic to a subset
of the real line and which is compact (i.e. closed and bounded),
perfect (i.e. containing no isolated points), and totally disconnected
(i.e. containing no intervals)\cite{hocking61}.  All Cantor sets are
homeomorphic (topologically equivalent) to each other
\cite{hocking61}, so when thinking about the topology of Cantor sets,
it is sufficient to imagine the middle-third Cantor set.  Cantor sets
may differ in their metric structure, that is, in the lengths and
separations of the segments that are deleted to form the set, and
different metric structures will in general result in different
fractal dimensions of the set.

We assume in the following that $\mathcal{L}_0$ is nowhere tangent to
the stable manifold; the results can be modified to account for such
tangencies, but it obfuscates our discussion.  The set $\mathbb{F}$ is
a Cantor set because it satisfies the three necessary requirements:
(1) It is closed and bounded, i.e. compact (by its definition.) (2) It
is totally disconnected, i.e. contains no intervals (because they have
been explicitly excluded.) (3) It is perfect, i.e. contains no
isolated points.  The third assertion follows from the Epistrophe
Theorem.  If there were an isolated surviving point, then directly on
its right there would have to be a (connected) open interval of
escaping points.  But any such interval must lie inside a single
escape segment.  Hence, the isolated point must be the left endpoint
of that escape segment and an intersection between $\mathcal{L}_0$ and
$\mathcal{S}$ (which is transverse by assumption.)  Therefore, the
point must have an epistrophe converging upon its left side, which
implies that the point is in fact not isolated.

The fractal $\mathbb{F}$ is the mutual boundary of the sets
$\mathbb{S}_T$ and $\mathbb{E}$.  Topologically, the boundary of a set
$U$ is the closure of $U$ (i.e. the smallest closed set containing
$U$) minus the interior of $U$ (i.e. the largest open set contained in
$U$.)  By definition, $\mathbb{F}$ is the boundary of $\mathbb{S}_T$.
We see that $\mathbb{F}$ is the boundary of $\mathbb{E}$ because any
point in $\mathbb{F}$ must have a point in $\mathbb{E}$ arbitrarily
close to it, otherwise it would be in the interior $\mathbb{S}$ of
$\mathbb{S}_T$.  We can now summarize the situation as follows: inside
the complex, the line of initial conditions is partitioned into three
sets: (1) the set $\mathbb{S}$ of all open surviving intervals, (2)
the set $\mathbb{E}$ of all open escaping intervals, and (3) their
mutual boundary $\mathbb{F}$; this boundary is a Cantor set.

Like all Cantor sets, $\mathbb{F}$ can be divided into two subsets
consisting of the \textit{accessible} and \textit{inaccessible}
points.  The fractal $\mathbb{F}$ is a line segment with a countable
number of open intervals removed, some being escape segments of
$\mathbb{E}$ and some being (generically) interiors of islands in
$\mathbb{S}$.  An endpoint of an escape segment is said to be
accessible from $\mathbb{E}$ because there exists a path beginning in
$\mathbb{E}$ and terminating on the endpoint without encountering any
other point outside of $\mathbb{E}$; we denote the countably infinite
set of points accessible from $\mathbb{E}$ by $\mathbb{A}_E$.
Similarly an endpoint of a segment in $\mathbb{S}$ is said to be
accessible from $\mathbb{S}$, and we denote the set of such points by
$\mathbb{A}_S$.  The set $\mathbb{A}_S$ may be countably infinite,
finite, or empty.  Together $\mathbb{A}_E$ and $\mathbb{A}_S$ form the
countably infinite set $\mathbb{A} = \mathbb{A}_E \cup \mathbb{A}_S$
of accessible points of $\mathbb{F}$ (i.e. the set of all endpoints of
deleted intervals, whether in $\mathbb{E}$ or $\mathbb{S}$.)  Since
the fractal $\mathbb{F}$ is uncountably infinite, most points in
$\mathbb{F}$ lie in the set, denoted $\mathbb{I}$, of inaccessible
points.

Every point of $\mathbb{A}_E$ lies on the stable manifold and hence has an
epistrophe converging upon it.  However, the points of $\mathbb{A}_S$
generically do not lie on the stable manifold, but are rather part of the
shoreline of $S$.  So we do not expect epistrophes to converge upon
the endpoints of a surviving segment of $\mathcal{L}_0$.  (This is
readily apparent in Fig.~\ref{fEPSC}.)  Nevertheless, the epistrophes
are dense in the fractal, or more precisely, the set $\mathbb{A}_E$ is
dense in $\mathbb{F}$. 

Because the escape segments exhibit what we called epistrophic
self-similarity in Sect.~\ref{sSelfSimilarity} and because
$\mathbb{F}$ is a Cantor set, we call $\mathbb{F}$ an {\em epistrophic
fractal}.

\section{Discussion and Future Research}

\label{sDiscussion}

The Epistrophe Theorem predicts the existence of an epistrophe
converging geometrically upon the endpoint of any escape segment; it
also characterizes the asymptotic behavior of the convergence in terms
of geometric quantities $\alpha$ and $\chi$.  However, none of our
present results say anything about how an epistrophe begins.  Most
notably, we do not predict the iterate at which an epistrophe starts,
and we certainly do not estimate the lengths or separation distances
of the early escape segments.

To understand the early behavior of an epistrophe, we must look at the
global topology of the tangle, which is the topic of the following
paper.  In that paper, we prove that another pattern appears in the
escape-time plots: the first segment $\epsilon_{k + \Delta}$ of an
epistrophe is spawned at some number of iterates $\Delta$ later than
the segment $\epsilon_k'$ upon which the epistrophe converges.  (This
fact was observed by Tiyapan and Jaff\'e \cite{Tiyapan93} and by Jung
and coworkers \cite{Ruckerl94a,Ruckerl94b,Lipp95} for certain lines of
initial conditions in scattering problems.)  This pattern is clearly
visible in Fig.~\ref{fEPSC} with $\Delta = 6$.  The pattern follows
from the fact that the topological structure of the map forces the
existence of a certain minimal required set of escape segments; this
minimal set can be shown to have the stated recursive pattern, for
$n_i$ sufficiently large.  Typically there are additional escape
segments (strophes) not predicted by the pattern; some of these
unpredicted segments are marked by asterisks in Fig.~\ref{fEPSC}.
Nevertheless, the minimal set seems to characterize the early
structure of the escape-time plots remarkably well.  However, based on
numerical evidence, we believe that for large enough iterate number,
the minimal set will eventually be overwhelmed by the unpredicted
segments, both in total number at a given iterate and in total
measure.  (Notice that the unpredicted segments in Fig.~\ref{fEPSC}
tend to be the longer segments at high iterate number.)

In future papers, we shall also present theorems concerning the
winding number (Fig.~\ref{fEPSC}), and we shall show how all these
concepts describe the ionization of a hydrogen atom placed in external
electric and magnetic fields.

We conclude by noting that standard references on
fractals~\cite{fractals} discuss at least three distinct types of
self-similarity: (1) regular self-similarity, as for the Cantor
middle-third set or the Koch snowflake; (2) asymptotic
self-similarity, which characterizes, for example, sequences of
period-doubling bifurcations; (3) statistical self-similarity, which
might describe coastlines or clouds.  To this list we add
``epistrophic self-similarity'', in which at all levels of resolution
there are asymptotically similar sequences, but additional unpredicted
segments may also persist.

\begin{acknowledgments}

The authors would like to thank Prof. Nahum Zobin for many useful
discussions.  This work was financially supported by the National
Science Foundation.

\end{acknowledgments}

\appendix

\section{A Family of Saddle-Center Maps}

\label{sSCTheory}

We define a map $\mathcal{M}(q_1,p_1) = (q_2,p_2)$ by 
\begin{subequations}
\begin{eqnarray}
q_2 & = & q_1 + {\partial
G(\bar{q},\bar{p}) \over \partial \bar{q}}, 
\label{r17} \\
p_2 & = & p_1 - {\partial G(\bar{q},\bar{p}) \over \partial
\bar{q}}, 
\label{r16} \\ 
\bar{q} & =
& (q_1 + q_2)/2, 
\label{r19} \\
\bar{p} & = & (p_1 + p_2)/2,
\label{r18}  
\end{eqnarray} 
\label{r23}
\end{subequations} 
where the ``Poincar\'e generator'' $G(q,p)$ is
\cite{Meyer70,Meyer92}
\begin{equation}
G(q,p) = \tau\left[ {{p}^2 \over 2m} +
V({q})\right] \label{r20} 
\end{equation}
and where
\begin{equation}
V({q}) = - \text{sech}(q) - f{q}
\label{r65}
\end{equation}
is a local potential well that goes to infinity on the left but only
has a potential barrier on the right.  It can be shown that this map
is canonical and that it is well-defined on the entire phase plane.

A fixed point of the map corresponds to a stationary point of
$G(q,p)$, where $\partial G/\partial p = \partial G/\partial q = 0$.
The fixed point is stable or unstable according to whether the Hessian
determinant $D = (\partial^2G/\partial p^2) \partial^2G/\partial q^2 -
(\partial^2G/\partial q\partial p)^2$ is positive or negative
\cite{Meyer70,Meyer92}.  Accordingly, the above map has exactly one
stable and one unstable fixed point, located at the local minimum and
the local maximum of $V(q)$.  The eigenvalues of an arbitrary fixed
point are given by
\begin{equation}
\alpha_{\pm} = {2 \pm \sqrt{-D} \over 2 \mp \sqrt{-D}}.
\label{r59}
\end{equation}
For the unstable fixed point, we have the explicit formula
\begin{subequations}
\begin{eqnarray}
D & = & - {\tau^2 \over m} \sqrt{ J (1-\sqrt{J}) \over 2}, \\
J & = & 1 - 4 f^2,
\end{eqnarray}
\label{r60}
\end{subequations}
from which the Liapunov factor $\alpha = \alpha_+>1$ can be computed.

\section{Proof of the Epistrophe Theorem}

\label{sEpistropheProof}

\subsection{Normal Form Near an \textsf{X}-Point}

\label{sVa}

We will need the following ``normal form'' theorem proved by
Moser\cite{Moser56}.

\begin{theorem}[Moser, Ref.~\cite{Moser56}]
In the neighborhood of an \textsf{X}-point of an analytic area- and
orientation-preserving map of the plane $(q,p) \mapsto (q',p')$, there
exists an analytic area-preserving change of coordinates $(q,p)
\mapsto (Q,P)$ that places the map into the normal form $(Q,P)
\mapsto (Q',P')$ satisfying
\begin{subequations}
\begin{eqnarray}
Q' & = & Q[\alpha + f(QP)], \\
P' & = & P/[\alpha + f(QP)], 
\end{eqnarray}
\label{r36}
\end{subequations}
where $\alpha$ is the Liapunov factor of the \textsf{X}-point.  The
coordinate change and the function $f(QP)$ are power series that have
a nonvanishing radius of convergence about the \textsf{X}-point.  The
function $f$ depends only on the product $Q$ times $P$, and $f(0) =
0$.
\label{t1}
\end{theorem}

A corollary of Moser's theorem is that the mapping (\ref{r36})
preserves the product
\begin{equation}
Q'P' = QP,
\end{equation}
so the hyperbolic curves $QP = constant$ are invariant sets under the
map.  We choose an open set $D$ that is convex in the $PQ$
coordinates, that contains the \textsf{X}-point of $\mathcal{M}$, and
in which the normal form converges.

\subsection{Convergence Factors Are Invariant Under Differentiable
Mappings}

Consider an infinite sequence of points $\mathbf{z}_n = (q_n,p_n)$
that lie on a differentiable curve $\mathcal{C}$ and that converge
geometrically to a point $\mathbf{z}_\infty \in \mathcal{C}$; i.e.,
the Euclidean arc length $d_n$, measured along $\mathcal{C}$ between
$\mathbf{z}_n$ and $\mathbf{z}_\infty$, satisfies
\begin{equation}
\lim_{n \rightarrow \infty} d_{n}\beta^n = K, 
\end{equation}
for some $\beta>1$ and $K > 0$.  We call $\beta$ the
\textit{convergence factor}.  For any map of the plane $\mathcal{N}$,
which is differentiable and locally invertible about
$\mathbf{z}_\infty$, define $D_n$ as the Euclidean arc length between
$\mathbf{Z}_n = \mathcal{N}(\mathbf{z}_n)$ and $\mathbf{Z}_\infty =
\mathcal{N}(\mathbf{z}_\infty)$, measured along
$\mathcal{N}(\mathcal{C})$.

\begin{lemma}
\label{t2}
The convergence factor $\beta$ is invariant under the mapping
$\mathcal{N}$, that is,
\begin{equation}
\lim_{n \rightarrow \infty} D_n \beta^n = J_0 K > 0.
\end{equation}
Here, $J_0 = | J\hat{\mathbf{t}} |$ where $J = \partial \mathcal{N} / \partial
\mathbf{z}|_{\mathbf{z}_\infty}$ is the $2\times 2$ Jacobian matrix evaluated
at $\mathbf{z} = \mathbf{z}_\infty$ and $\hat{\mathbf{t}}$ is the unit tangent to
the curve $\mathcal{C}$ at $\mathbf{z}_\infty$.
\end{lemma}
The notation $| \; \; |$ denotes the standard Euclidean vector norm.
The proof of this lemma is a simple exercise.

\subsection{Iterates of a Curve Intersecting the Stable Manifold
Approach the Unstable Manifold}

Consider any differentiable curve $\mathcal{C}_0$ having a transverse
intersection with the stable manifold $\mathcal{S}$ of $\mathcal{M}$
at a point $\mathbf{r}_0 = (q_0,p_0)$ having canonical length $r_0 =
s(\mathbf{r}_0)$ along $\mathcal{S}$; we assume $\mathcal{C}_0$ does
not intersect the segment of $\mathcal{S}$ joining
$\mathbf{z}_\textsf{X}$ to $\mathbf{r}_0$ (Fig.~\ref{fT4}).  The
differentiable curve $\mathcal{C}_1 = \mathcal{M}(\mathcal{C}_0)$
intersects $\mathcal{S}$ at a point $\mathbf{r}_1 = (q_1,p_1) =
\mathcal{M}(\mathbf{r}_0)$, which is closer to the $\textsf{X}$-point
$\mathbf{z}_\textsf{X}$.  By Eq.~(\ref{r42}) $\mathbf{r}_1$ has
canonical length $r_1 = s(\mathbf{r}_1) = r_0 / \alpha$.  Applying
$\mathcal{M}$ repeatedly, we obtain an infinite sequence of curves
$\mathcal{C}_n$ that intersect $\mathcal{S}$ in a sequence
$\mathbf{r}_n$ converging geometrically to $\mathbf{z}_\textsf{X}$
with convergence factor equal to the Liapunov factor $\alpha$:
$r_n\alpha^n = r_0 \ne 0$.

The curves $\mathcal{C}_n$ ``approach the unstable manifold'' in the
sense that every point $\mathbf{z}_{\mathcal{U}}$ on $\mathcal{U}$ is
the limit of a sequence of points $\mathbf{z}_n \in \mathcal{C}_n$
converging geometrically with factor $\alpha$.  

\begin{lemma}
For any differentiable curve $\mathcal{C}_0$ passing transversely
through $\mathcal{S}$ at a point $\mathbf{r}_0$ and for any
differentiable curve $\bar{\mathcal{C}}$ passing transversely through
$\mathcal{U}$ at a point $\mathbf{z}_{\mathcal{U}}$, the curves
$\mathcal{C}_n = \mathcal{M}^n(\mathcal{C}_0)$ and $\bar{\mathcal{C}}$
intersect for all $n$ large enough.  We further assume $\mathcal{C}_0$
does not intersect the segment of $\mathcal{S}$ joining
$\mathbf{z}_\textsf{X}$ to $\mathbf{r}_0$.  We then choose
$\mathbf{z}_n$ to be the point in $\mathcal{C}_n \cap
\bar{\mathcal{C}}$ closest to $\mathbf{z}_{\mathcal{U}}$ (as measured
along $\bar{\mathcal{C}}$.)  The sequence $\mathbf{z}_n$ satisfies
\begin{eqnarray} 
\lim_{n \rightarrow \infty} \mathbf{z}_n & = & \mathbf{z}_{\mathcal{U}}, \\ 
\lim_{n \rightarrow \infty} |\mathbf{z}_{n} - \mathbf{z}_{\mathcal{U}}| \alpha^n
 & = &  A(\mathbf{z}_{\mathcal{U}}, \hat{\mathbf{t}}) r_0  \ne 0, 
\label{r28}
\end{eqnarray}
where $\alpha$ is the Liapunov factor of $\mathbf{z}_\textsf{X}$ and
$A(\mathbf{z}_{\mathcal{U}}, \hat{\mathbf{t}})$ is a function that,
for a given map $\mathcal{M}$, depends only on the intersection point
$\mathbf{z}_{\mathcal{U}}$ and the tangent $\hat{\mathbf{t}}$ to the
curve $\bar{\mathcal{C}}$ at $\mathbf{z}_{\mathcal{U}}$.
\label{t4}
\end{lemma}

\noindent \textit{Proof:} First suppose $\mathbf{z}_{\mathcal{U}}$ and
$\mathbf{r}_0$ lie within the domain $D$ in which the normal form
applies (Sect.~\ref{sVa}).  Within $D$, we use the normal-form
coordinates $(Q,P)$, in which case $\mathbf{z}_{\mathcal{U}}$ and
$\mathbf{r}_0$ are represented by $\mathbf{Z}_{\mathcal{U}} =
(Q_{\mathcal{U}},0)$ and $\mathbf{R}_0 = (0, R_0)$.  By
Lemma~\ref{t2}, if we prove Lemma~\ref{t4} in the $(Q,P)$ coordinates
then we have proved Lemma~\ref{t4} in the original $(q,p)$
coordinates.

The point $\mathbf{Z}_n = (Q_n, P_n)$ in $\mathcal{C}_n \cap
\bar{\mathcal{C}}$ is the $n$th iterate of some point $\mathbf{Z}_n^0 =
(Q_n^0, P_n^0)$ in $\mathcal{C}_0$.  (See Fig.~\ref{fT4}.)  Thus,
iterating Eqs.~(\ref{r36}), we see that
\begin{subequations}
\begin{eqnarray}
Q_n & = & Q_n^0[ \alpha + f(Q_n P_n)]^n, \\
P_n & = & P_n^0[ \alpha + f(Q_n P_n)]^{-n}.
\end{eqnarray}
\label{r37}
\end{subequations}
Near the point $\mathbf{R}_0$ the curve $\mathcal{C}_0$ is the graph
of a differentiable function $P = C_0(Q)$, and similarly, near the
point $\mathbf{Z}_{\mathcal{U}}$, the curve $\bar{\mathcal{C}}$ is the
graph of a differentiable function $Q = \bar{C}(P)$.  Combining this
with Eqs.~(\ref{r37}), we find that for $n$ large enough $P_n$ must
satisfy
\begin{subequations}
\begin{eqnarray}
\bar{C}(P_n) & = & Q_n^0[ \alpha + f(P_n\bar{C}(P_n))]^n, 
\label{r34}\\
P_n & = & C_0(Q_n^0)[ \alpha + f(P_n\bar{C}(P_n))]^{-n}.
\label{r35}
\end{eqnarray}
\end{subequations}
Solving for $Q_n^0$ in Eq.~(\ref{r34}) and inserting the result into
Eq.~(\ref{r35}), we find
\begin{equation}
P_n = G_n(P_n) [ \alpha + F(P_n) ]^{-n},
\label{r38}
\end{equation}
where 
\begin{eqnarray}
G_n(P) & = & C_0\mathbf{(} \bar{C}(P) [\alpha + F(P) ]^{-n}\mathbf{)}, \label{r39}\\
F(P) & = & f\mathbf{(}P \bar{C}(P)\mathbf{)}. 
\end{eqnarray}
The functions $G_n(P)$ and $F(P)$ have the following properties: (1)
$G_n(P)$ and $F(P)$ are well-defined, differentiable functions in the
neighborhood of $P=0$ (for $n$ sufficiently large); (2) $F(0) = 0$;
(3) $G_n(0) \ne 0$.  Observe that Eq.~(\ref{r38}) is an implicit
expression for $P_n$.

\noindent \textit{Lemma 2a: For each $n$, choose $P_n>0$ to be the
smallest real solution to Eq.~(\ref{r38}).  Then for $n$ sufficiently
large, $P_n$ is well-defined and $\lim_{n \rightarrow \infty} P_n
\alpha^n = R_0$.}

\noindent \textit{Proof:} Define $E_n(P) = P[\alpha + F(P)]^n / G_n(P)
- 1$.  Then $P_n$ satisfies Eq.~(\ref{r38}) in the neighborhood of $0$
if and only if it is a zero of $E_n$.  Let $\epsilon > 0$ be given.
Recalling that $F(0) = 0$ and $\alpha>1$, we assume that $\epsilon$ is
small enough so that $\alpha + F(\epsilon) > 1$ and
$\bar{C}(\epsilon)$ is well-defined.  This implies
\begin{equation}
\lim_{n \rightarrow \infty} G_n(\epsilon) = C_0(0) = R_0
> 0, 
\label{r7}
\end{equation}
and furthermore,
\begin{equation}
\lim_{n \rightarrow \infty} E_n(\epsilon) = \lim_{n \rightarrow
\infty} {\epsilon[\alpha + F(\epsilon)]^n \over R_0} = +\infty.
\end{equation}
Combining this result with the fact that $E_n(0) = -1$ [since $G_n(0)
\ne 0$], we see that there must be a zero of $E_n$ between $0$ and
$\epsilon$ for all $n$ large enough.  Since $\epsilon > 0$ was
arbitrarily small, we can construct a sequence $P_n$ of zeros of
$E_n(P)$ such that $P_n$ goes to $0$.  We choose the sequence $P_n$ to
consist of the smallest positive roots of $E_n$.

Next, notice that
\begin{eqnarray}
\lim_{n \rightarrow \infty} nP_n & = & \lim_{n \rightarrow \infty} n[\alpha+F(P_n)]^{-n}G_n(P_n) \nonumber \\
& = & \lim_{n \rightarrow \infty} n[\alpha+F(P_n)]^{-n} R_0
= 0,
\label{r40}
\end{eqnarray}
where the first equality follows from Eq.~(\ref{r38}), the second from
Eq.~(\ref{r7}), and the last from $F(0) = 0$.  Finally,
\begin{eqnarray}
\lim_{n \rightarrow \infty} \ln (P_n\alpha^n)
& = & \ln R_0 - \lim_{n \rightarrow \infty} n \ln [1 + F(P_n)/\alpha] \nonumber \\
& = & \ln R_0 -  {1 \over \alpha} \left.{dF \over dP}\right|_{P=0} 
\lim_{n \rightarrow \infty} n P_n \nonumber \\
& = & \ln R_0,
\label{r61}
\end{eqnarray}
where the first equality follows from Eqs.~(\ref{r38}) and (\ref{r7}),
the second by expanding $F(P_n)$ about $P_n = 0$, and the last from
Eq.~(\ref{r40}).  This completes the proof of Lemma 2a.

\vskip 12pt

The sequence $P_n$ yields the sequence of points $\mathbf{Z}_n =
(Q_n,P_n)= (\bar{C}(P_n),P_n)$ in the statement of Lemma~\ref{t4}.
Equation~(\ref{r28}) in the normal-form coordinates follows
immediately from Lemma 2a
\begin{eqnarray}
\lim_{n \rightarrow \infty} |\mathbf{Z}_{n} - \mathbf{Z}_{\mathcal{U}}| \alpha^n& = & \lim_{n \rightarrow \infty}[ P_{n}^2 + (Q_{n} - Q_{\mathcal{U}})^2]^{1/2}\alpha^n \nonumber \\
& = & A(\hat{\mathbf{t}})R_0 \ne 0,  
\end{eqnarray}
where $A(\hat{\mathbf{t}}) = [ 1 + (d \bar{C} / d P|_{P = 0})^2]^{1/2}$
obviously depends only on the tangent $\hat{\mathbf{t}}$ to
$\bar{\mathcal{C}}$ at the intersection point $\mathbf{Z}_{\mathcal{U}} =
(Q_{\mathcal{U}}, 0)$.  Lemma~\ref{t2} shows that the transformation
back to the original $(q,p)$ coordinates introduces a Jacobian factor
(dependent on $\mathbf{z}_{\mathcal{U}}$ and $\hat{\mathbf{t}}$) which can be
absorbed into a new $A(\mathbf{z}_{\mathcal{U}}, \hat{\mathbf{t}})$.
 
At this point Lemma \ref{t4} is proved provided $\mathbf{z}_{\mathcal{U}}$
and $\mathbf{r}_0$ lie within $D$, where the normal form is valid.  To
extend the theorem to arbitrary $\mathbf{r}_0$ on the stable manifold and
arbitrary $\mathbf{z}_{\mathcal{U}}$ on the unstable manifold, we use the
following ``forward-backward mapping trick.''

First suppose $\mathbf{z}_{\mathcal{U}}$ lies within ${D}$ but
$\mathbf{r}_0$ lies outside of ${D}$.  We apply $\mathcal{M}$ a finite
number of times $k$ until $\mathbf{r}_k$ lies within $D$, and then we
repeat the above argument.

On the other hand, if $\mathbf{z}_{\mathcal{U}}$ lies outside of $D$
we apply the inverse map $\mathcal{M}^{ -1}$ to
$\mathbf{z}_{\mathcal{U}}$ $j$ times until
$\mathbf{z}^{-j}_{\mathcal{U}}
=\mathcal{M}^{-j}(\mathbf{z}_{\mathcal{U}})$ lies within ${D}$.  Then
by the preceding argument, we identify the sequence of points
$\mathbf{z}_{n-j}^{-j} \in \mathcal{C}_{n-j} \cap
\mathcal{M}^{-j}(\bar{\mathcal{C}})$ converging to
$\mathbf{z}^{-j}_{\mathcal{U}}$.  We map this sequence forward $j$
times to arrive at the sequence $\mathbf{z}_n \in \mathcal{C}_{n} \cap
\bar{\mathcal{C}}$ converging to $\mathbf{z}_{\mathcal{U}}$.  By
Lemma \ref{t2}, the convergence factor remains $\alpha$; we also
acquire a Jacobian factor $J_0$ from the transformation, but it will
depend only on $\mathbf{z}_{\mathcal{U}}$ and $\hat{\mathbf{t}}$ and
can be absorbed into a new $A(\mathbf{z}_{\mathcal{U}},
\hat{\mathbf{t}})$.  $\mathcal{QED}$

\vskip 12pt

Note that $A(\mathbf{z}_{\mathcal{U}}, \hat{\mathbf{t}})$ in
Lemma~\ref{t4} does \textit{not} depend on the curve $\mathcal{C}_0$.
The following corollary exploits this fact.

\begin{lemma}

For any two differentiable curves $\mathcal{C}_0$ and
$\mathcal{C}_0'$, each passing transversely through $\mathcal{S}$ at
the points $\mathbf{r}_0$ and $\mathbf{r}_0' \ne \mathbf{r}_0$,
respectively, and for any differentiable curve $\bar{\mathcal{C}}$
passing transversely through $\mathcal{U}$ at a point
$\mathbf{z}_{\mathcal{U}}$, the curves $\mathcal{C}_n =
\mathcal{M}^n(\mathcal{C}_0)$ and $\mathcal{C}'_n=
\mathcal{M}^n(\mathcal{C}_0')$ both intersect $\bar{\mathcal{C}}$ for
all $n$ large enough.  We further assume $\mathcal{C}_0$ (or
$\mathcal{C}_0'$) does not intersect the segment of $\mathcal{S}$
joining $\mathbf{z}_\textsf{X}$ to $\mathbf{r}_0$ (or
$\mathbf{r}_0'$).  We choose $\mathbf{z}_n$ (or $\mathbf{z}'_n$) to be
the point in $\mathcal{C}_n \cap \bar{\mathcal{C}}$ (or
$\mathcal{C}'_n \cap \bar{\mathcal{C}}$) closest to
$\mathbf{z}_{\mathcal{U}}$ (as measured along $\bar{\mathcal{C}}$.)
The sequences $\mathbf{z}_n$ and $\mathbf{z}'_n$ satisfy
\begin{eqnarray} 
\lim_{n \rightarrow \infty} \mathbf{z}_n 
= \lim_{n \rightarrow \infty} \mathbf{z}'_n 
& = & \mathbf{z}_{\mathcal{U}}, \\ 
\lim_{n \rightarrow \infty} |\mathbf{z}_{n} - \mathbf{z}'_{n}| \alpha^n
 & = &  A(\mathbf{z}_{\mathcal{U}}, \hat{\mathbf{t}}) d_0  \ne 0, 
\end{eqnarray}
where $d_0$ is the canonical length between $\mathbf{r}_0$ and
$\mathbf{r}_0'$ (measured along $\mathcal{S}$), $\alpha$ is the
Liapunov factor of $\mathbf{z}_\textsf{X}$, and
$A(\mathbf{z}_{\mathcal{U}}, \hat{\mathbf{t}})$ is the same function
as in Lemma \ref{t4}.
\label{t3}
\end{lemma}

\subsection{The Epistrophe Theorem}

We now complete the proof of the Epistrophe Theorem.  We cut the
$\mathcal{U}$-boundary of the capture zone $C_1$ (see Fig.~\ref{fX1})
at an arbitrary point $\mathbf{z}_{cut}$, creating the following two
curves: (1) $\mathcal{C}_{\mathbf{Q}_0}$ begins at $\mathbf{z}_{cut}$
and continues backward along $\mathcal{U}$ until just past
$\mathbf{Q}_0$; (2) $\mathcal{C}_{\mathbf{P}_1}$ begins at
$\mathbf{z}_{cut}$ and continues forward along $\mathcal{U}$ until
just past $\mathbf{P}_1$.  Applying Lemma~\ref{t3} to $\mathcal{C}_0 =
\mathcal{C}_{\mathbf{Q}_0}$, $\mathcal{C}_0' =
\mathcal{C}_{\mathbf{P}_1}$, and $\bar{\mathcal{C}} = \mathcal{L}_0$
we find the following.

\begin{lemma}
For any differentiable curve $\mathcal{L}_0$ passing transversely
through $\mathcal{U}$ at a point $\mathbf{z}_{\mathcal{U}}$, the
capture zones $C_n$ intersect $\mathcal{L}_0$ for all $n$ large
enough.  We choose $\epsilon_n$ to be the connected component of $C_n
\cap \mathcal{L}_0$ closest to $\mathbf{z}_{\mathcal{U}}$ (as
measured along $\mathcal{L}_0$.)  The $\epsilon_n$ converge upon
$\mathbf{z}_{\mathcal{U}}$ with the Euclidean length $\lambda_n$ of
$\epsilon_n$ satisfying
\begin{equation}              
\lim_{n \rightarrow \infty} \lambda_n \alpha^n 
= A(\mathbf{z}_{\mathcal{U}}, \hat{\mathbf{t}}) s(C_1)
\ne 0, 
\label{r52}
\end{equation} 
\label{t6}
where $s(C_1)$ is the canonical length of the $\mathcal{S}$-boundary
of $C_1$ [Eq.~(\ref{r11})].
\end{lemma}
So long as $\rho(\mathbf{z}_{\mathcal{U}}) \ne 0$, Lemmas~\ref{t4} --
\ref{t6} hold when the Euclidean length $\lambda$ is replaced by the
measure $\mu$ defined by Eq.~(\ref{r46}), except that
$A(\mathbf{z}_{\mathcal{U}}, \hat{\mathbf{t}})$ is multiplied by
$\rho(\mathbf{z}_{\mathcal{U}})$.  In particular, Eq.~(\ref{r52})
yields Eq.~(\ref{r47})
\begin{equation} 
\lim_{n \rightarrow \infty} \mu_n \alpha^n =
[A(\mathbf{z}_{\mathcal{U}}, \hat{\mathbf{t}})
\rho(\mathbf{z}_{\mathcal{U}})] s(C_1) \equiv K_\mu \ne
0,
\label{r53}
\end{equation}
where $\mu_n$ is the measure of $\epsilon_n$.

Between $\epsilon_n$ and $\epsilon_{n+1}$ lies a gap on
$\mathcal{L}_0$ with measure $\gamma_n$.  Applying Lemma~\ref{t3} to
$\mathcal{C}_0 = \mathcal{C}_{\mathbf{P}_1}$ and $\mathcal{C}_0' =
\mathcal{M}(\mathcal{C}_{\mathbf{Q}_0})$ yields Eq.~(\ref{r48})
\begin{equation} 
\lim_{n \rightarrow \infty} \gamma_n \alpha^n 
= [A(\mathbf{z}_{\mathcal{U}}, \hat{\mathbf{t}}) \rho(\mathbf{z}_{\mathcal{U}})] s(E_1) 
\equiv K_\gamma \ne 0,
\end{equation}
from which also follows Eq.~(\ref{r50})
\begin{equation}
\lim_{n \rightarrow \infty} {\mu_n \over \gamma_n} 
= {s(C_1) \over s(E_1)}
= {s(C_0) \over s(E_0)}
= {K_\mu \over K_\gamma} \equiv \chi \ne 0.
\label{r54}
\end{equation}
Similarly, Lemma~\ref{t4} applied to $\mathcal{C}_0 =
\mathcal{C}_{\mathbf{P}_1}$ implies Eq.~(\ref{r49})
\begin{equation} 
\lim_{n \rightarrow \infty} \delta_n \alpha^n 
= [A(\mathbf{z}_{\mathcal{U}}, \hat{\mathbf{t}}) \rho(\mathbf{z}_{\mathcal{U}})] s(\mathbf{P}_1) 
\equiv K_\delta \ne 0,
\label{r63}
\end{equation}
where $\delta_n$ is the measure between $\epsilon_n$ and
$\mathbf{z}_{\mathcal{U}}$.  From Eq.~(\ref{r63}) follows Eq.~(\ref{r57})
\begin{equation}
\lim_{n \rightarrow \infty} {\mu_n \over \delta_n} 
= {s(C_1) \over s(\mathbf{P}_1)}
= {s(C_0) \over s(\mathbf{P}_0)}
= {K_\mu \over K_\delta} \equiv \phi \ne 0.
\label{r62}
\end{equation}

The Epistrophe Theorem requires studying intersections of the
pre-iterates of $E_0$ with $\mathcal{L}_0$.  In the results above, we
have studied intersections of the forward iterates of $C_0$ with
$\mathcal{L}_0$.  One may, of course, translate between these two
viewpoints by replacing $\mathcal{M}$ with $\mathcal{M}^{-1}$.

The final equality for $\phi$ follows from
\begin{equation}
u(\mathbf{P}_0) 
= \sum_{n = 0}^\infty 
[ u(C_0) + u(E_0)\alpha^{-1}] \alpha^{-n} 
= u(C_0) { \alpha + \chi \over \alpha - 1}.
\end{equation}

\begin{figure}
\caption{\label{fSOSsc} A phase space portrait is shown for a map
possessing a single homoclinic tangle.  [The map is defined by
Eqs.~(\ref{r23}) -- (\ref{r65}) with $\tau = 1.5$, $f = 0.25$, and $m
= 0.57$.]  The \textsf{X}-point $\mathbf{z}_\mathsf{X}$ has a stable
manifold $\mathcal{S}$ and unstable manifold $\mathcal{U}$ which cross
repeatedly to form the tangle.  The primary intersection point
$\mathbf{P}_0$ defines the complex, the northern and southern
boundaries of which are the segments of $\mathcal{S}$ and
$\mathcal{U}$ joining $\mathbf{P}_0$ to $\mathbf{z}_\textsf{X}$.
Orbits escape the complex by mapping from $E_{-1}$ into $E_{0}$ and
then move away through successive iterates: $E_1$, $E_2$, $E_3$,
... .  Orbits are captured by mapping from $C_0$ into $C_1$.  The line
of initial conditions $\mathcal{L}_0$ coincides with $q = 0$.}
\end{figure}

\begin{figure}
\caption{\label{fEPSC} Escape data $n_i$ and $n_w$ are plotted for the
map in Fig.~\ref{fSOSsc}.  Shown on the right is the number of
iterates $n_i$ required for a point to escape from the complex; it is
plotted as a function of $p$ parameterizing the line of initial
conditions $\mathcal{L}_0$.  Several sequences (epistrophes) of escape
segments are indicated with bold arrows.  Several escape segments
(strophes) are marked by asterisks; these segments are not easily
predicted from the current level of theory.  Plotted on the left is
the winding number of the trajectory as it escapes to infinity.}
\end{figure}

\begin{figure}
\caption{\label{fX1} A qualitative representation of a homoclinic
tangle satisfying Assumptions 1 -- 5.  It displays features similar to
the data in Fig.~\ref{fSOSsc}, but the escape zones and capture zones
are more clearly labeled.}
\end{figure}

\begin{figure}
\caption{\label{flinear} Introducing new canonical coordinates
$(Q,P)$, the saddle-center map $\mathcal{M}$, linearized about the
\textsf{X}-point $\mathbf{z}_\textsf{X}$, has the simple form $Q
\mapsto \alpha Q$, $P \mapsto (1/\alpha)P$.  The stable manifold
$\mathcal{S}$ and unstable manifold $\mathcal{U}$ coincide
respectively with the positive $P$ and negative $Q$ axes.  A sequence
of backward iterates of the escape zone $E_{-k}$ is shown shaded.
Under each iterate, the width of the lobe decreases by $1/\alpha$
while the height increases by $\alpha$.  Eventually, these iterates
must intersect the line of initial conditions $\mathcal{L}_0$, shown
as a horizontal line passing through the $P$ axis.  The escape
segments $\epsilon_{k+1}$, $\epsilon_{k+2}$, $...$ which are created
by these intersections decrease in size by the same geometric factor
$\alpha$ (in the asymptotic limit).  They eventually converge upon the
intersection $\mathbf{z}_{\mathcal{S}}$ of $\mathcal{L}_0$ with
$\mathcal{S}$.  }
\end{figure}

\begin{figure}
\caption{\label{fmu} The length $\mu_n$ of an escape segment is
plotted as a function of iterate number $n = n_i$ for the five
epistrophes \textsf{a} -- \textsf{e} shown in Fig.~\ref{fEPSC}.  The
lines all have the same slope, equal to $-\log \alpha$, where $\alpha
= 2.776$ is the Liapunov factor of the \textsf{X}-point in
Fig.~\ref{fSOSsc}.  This factor can be computed analytically from
Eqs.~(\ref{r59}) and (\ref{r60}).  Clearly as $n \rightarrow \infty$,
$\mu_n$ decays geometrically with $\mu_n \rightarrow (constant) \times
\alpha^{-n}$.}
\end{figure}

\begin{figure}
\caption{\label{fgamma}
Analogous to Fig.~\ref{fmu}, the distance $\gamma_n$
between successive escape segments is plotted as a function of iterate
number $n = n_i$.  It is again readily apparent that as $n \rightarrow
\infty$, $\gamma_n \rightarrow (constant) \times \alpha^{-n}$, where
$\alpha = 2.776$ is the same as in Fig.~\ref{fmu}.}
\end{figure}

\begin{figure}
\caption{\label{fchi} Plotted is the ratio $\mu_n/\gamma_n$ of the
data in Fig.~\ref{fmu} to the data in Fig.~\ref{fgamma}.  The ratio
for each epistrophe \textsf{a} -- \textsf{e} has the same asymptotic
value $\chi = 2.2113$, which agrees with an independent computation of
$\chi$ from the formula $\chi =
u(E_0)/u(C_0) = \lim_{k \rightarrow
\infty} |\mathbf{Q}_{-k} - \mathbf{P}_{-k}|/|\mathbf{P}_{-k} - \mathbf{Q}_{-(k+1)}|$.}
\end{figure}

\begin{figure}
\caption{\label{fsets} The above schematic illustrates the relations
between various subsets of $\mathcal{L}_0$ which are important for
understanding the fractal nature of escape.  Each branching of the
tree represents a partition of the upper set into the two lower
disjoint subsets.}
\end{figure}

\begin{figure}
\caption{\label{fT4} This diagram, used for proving Lemma~\ref{t4},
depicts the neighborhood of $\mathbf{z}_\textsf{X}$ (the origin) in
normal-form coordinates $(Q, P)$.  The dashed lines are the invariant
hyperbolas of the map.  The curves $\mathcal{C}_n$,
$\mathcal{C}_{n+1}$, and $\mathcal{C}_{n+2}$ are the $n$th, $(n+1)$th,
and $(n+2)$th iterates of $\mathcal{C}_0$, respectively.  Similarly,
the points $\mathbf{Z}_n$, $\mathbf{Z}_{n+1}$, and $\mathbf{Z}_{n+2}$
lying on $\bar{\mathcal{C}}$ are the $n$th, $(n+1)$th, and $(n+2)$th
iterates of the points $\mathbf{Z}^0_n$, $\mathbf{Z}^0_{n+1}$, and
$\mathbf{Z}^0_{n+2}$, respectively.  The unlabeled dots represent
intermediate iterates of these points.}
\end{figure}

\printfigures


\begin{thebibliography}{99}
\bibitem{footnote1} We also have in mind applications to the
semi-classical propagation of quantum wave functions, for which it is
natural to consider an initial curve (Lagrangian manifold) in phase
space \cite{Maslov}.  Lines of initial conditions are also common in
studies of classical chaotic scattering (as in
Refs.~\cite{Jung87,Ruckerl94a,Ruckerl94b,Lipp95,Jung99,Butikofer00}.)
\bibitem{footnote3} An endpoint could also lie on $\mathcal{U}$, in
which case it would be on the boundary of the complex and no
epistrophe would converge upon it.  See Fig.~\ref{fSOSsc}.
\bibitem{footnote4} Note that $P$ and $Q$ are canonical coordinates, whereas
$\mathbf{P}_n$ and $\mathbf{Q}_n$ are points in the phase plane.
\bibitem{footnote5} In exceptional circumstances, there may be other
intervals that survive.  For example, it could happen that
$\mathcal{L}_0$ coincides with the stable manifold over an interval of
nonzero length, a nongeneric situation.
\bibitem{footnote6} Both $\mathbb{S}_T$ and $\mathbb{E}$ are subsets
of both the line of initial conditions and the phase plane, but for
topological purposes, we view them as subsets of $\mathcal{L}_0$.
That is, an open interval of $\mathcal{L}_0$ is viewed as an open set,
even though it is not an open subset of the phase plane.
\end{thebibliography}
\end{document}